\documentclass[twocolumn]{aastex701}
\usepackage{amsmath, amssymb, graphics}
\usepackage{graphicx}
\usepackage{caption}
\def\cc {\ifmmode{{\rm cm}^{-2}}\else{${\rm
cm}^{-2}$}\fi}
\def\ccc {\ifmmode{{\rm cm}^{-3}}\else{${\rm
cm}^{-3}$}\fi}

\def\kms  {km\,s$^{-1}$}
\def\Lkkmspc  {K\,km\,s$^{-1}$\,pc$^{2}$}
\def\aco {\ifmmode{^{12}{\rm CO}(J=1-0)}\else{$^{12}{\rm
CO}(J=1-0)$}\fi}
\def\bco {\ifmmode{^{12}{\rm CO}(J=2-1)}\else{$^{12}{\rm
CO}(J=2-1)$}\fi}
\def\cco {\ifmmode{^{12}{\rm CO}(J=3-1)}\else{$^{12}{\rm
CO}(J=3-2)$}\fi}

\def\ahcn {{${\rm
HCN}(J=1-0)$}}

\def\mh     {H$_{2}$}

\def\Msun{\ifmmode{{\rm M}_{\odot}}\else{{M$_{\odot}$}}\fi}
\def\deg {$^{\circ}$}

\def\arcsec {$^{\prime\prime}$}

\newcommand\sgra {Sgr~A$^*$}

\usepackage{xcolor}

\begin{document}

\title{The CO--to--\mh\ conversion factor in the Milky Way's central parsec}

\author[orcid=0000-0001-9300-354X,sname='Mark Gorski']{Mark D. Gorski}
\affiliation{Center for Interdisciplinary Exploration and Research in Astrophysics, Northwestern University}
\email[show]{mark.gorski@northwestern.edu}  

\author[orcid=0000-0001-8986-5403, sname='Elena Murchikova']{Lena Murchikova} 
\affiliation{Center for Interdisciplinary Exploration and Research in Astrophysics, Northwestern University}
\affiliation{Department of Physics and Astronomy, Northwestern University}
\affiliation{School of Natural Sciences, Institute for Advanced Study, Princeton}
\email{lena@northwestern.edu}

\begin{abstract}


Carbon monoxide (CO) emission is a widely used tracer of molecular hydrogen (\mh) in the interstellar medium (ISM), owing to its abundance, low excitation energy, and ease of detection in cold molecular environments, in contrast to \mh \, itself.
While the CO-to-\mh \, conversion factor is often assumed to be constant across the disks of galaxies, deviations are observed in extreme environments such as the central molecular zone (CMZ) in galactic nuclei.
Here we present the first estimate of the CO-to-\mh\ conversion factor on sub-kpc scales.
We calculate CO-to-\mh\ conversion in the Milky Way's Circumnuclear Disk/Ring (CND/CNR) at $\sim 1$ pc radius around the Galactic Center black hole.
{
We derive a conversion factor of 
$\alpha_\mathrm{CO} \simeq 4.5\pm2.5$~\Msun(\Lkkmspc)$^{-1}$ 
or 
X[CO] $\simeq (2.1\pm1.1)\times 10^{20}$~cm$^{-2} (\mathrm{K \, km \, s^{-1}})^{-1}$.
This value is consistent with the Galactic disk but higher than CMZ.
}

\end{abstract}


\keywords{\uat{Galaxies}{573} ---  \uat{Interstellar medium}{847} --- \uat{Interstellar molecules}{849} --- \uat{Radio spectroscopy}{1359} ---\uat{Galactic Center}{565} --- \uat{Galaxy circumnuclear disk}{581}}


\section{Introduction}\label{sec:intro}%

Carbon monoxide (CO) is the most commonly used tracer of the molecular gas in the interstellar medium (ISM) \citep{Bolatto2013b}.
It has a relatively high abundance in the molecular ISM, a low excitation energy ($\sim5$~K), a low critical density, and can be excited even in cold $\sim10$~K ISM. 
In contrast, molecular hydrogen (\mh), while much more common, has its lowest energy transition is in the far-infrared at $\sim500$~K. 
As a result, estimates of the total molecular gas mass in both local and distant galaxies and their substructures often rely on the CO--to--\mh\ conversion factor.

The CO--to--\mh\ conversion factor is commonly denoted as either a scaling relation between CO luminosity measured in \Lkkmspc \, and the mass of molecular gas measured in \Msun \, ($\alpha_\mathrm{CO}$), or 
a scaling relation between the integrated CO intensity measured in $\mathrm{K} \,\mathrm{km \,s}^{-1}$ and the H$_2$ column density measured in $\mathrm{cm}^{-2}$ ($X_\mathrm{CO}$).

The CO--to--\mh\ conversion factor is often assumed to be constant throughout galactic disks.
Several have shown this to be generally true regardless of the CO abundance or radiative transfer effects \citep{Sandstrom2013,Bolatto2013b}.
However, in some environments a deviation from the galactic mean is expected, e.g., high-density, low-metallicity, high-temperature, and increased velocity dispersion environments. 
For example, in the Milky Way's central molecular zone (CMZ), located in the inner 300 pc of the Galactic center, the CO--to--\mh\ is a factor of $3-10$ lower than the galactic disk \citep{Kohno2024,Oka1998}.

In this work we used observations conducted with the Atacama Large Millimeter/submillimeter Array (ALMA) of a section of the Galactic Center Circumnuclear Disk/Ring (CND/CNR) in the \bco\ line at 230.538 GHz.
These are the first observations of this region in this line since \citealt{Marshall1994}.
{Combined with known properties of the CND deduced from observations in a multitude of lines \citep{James2021, Hsieh2018, Tsuboi2018, Requena-Torres2012, Lau2013, Liu2013,Oka2011, Etxaluze2011}, we deduce the CO--to--\mh\ conversion factor in the Milky Way's CND without relying on the traditional assumption of virialization. }

This paper is organized as follows. In Section \ref{sec:observations} we describe the observations and summarize the data processing techniques. In Section \ref{sec:co2mh} we calculate CO--to--\mh\ conversion factor in the Milky Way's CND in two different ways and discuss the uncertainties. In Section~\ref{sec:assume} we summarize the results and give the final estimates for $\alpha_\textrm{CO}$ and $X_\mathrm{CO}$.

\begin{figure}
    \centering
    \includegraphics[width=\linewidth]{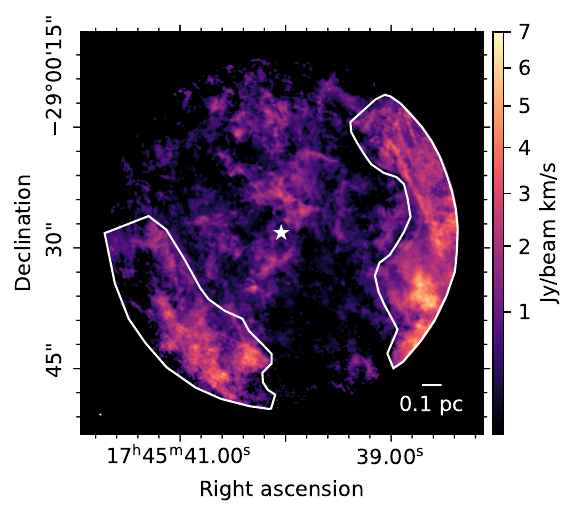}
    \caption{
    Integrated flux map of the \bco\ emission around \sgra.
    The map is primary beam corrected.
    The calculations of the CO--to--\mh\ conversion factor use the region outlined with the white contours.
    The position of \sgra\ is marked with a white star.
    }
    \label{fig:CNDreg}
\end{figure}

\section{Observations} \label{sec:observations}
\subsection{Calibration and Image quality}
We combined data from ALMA Band 6 observations of the Milky Way's central black hole \sgra\ from projects 2016.1.00870.S, 2017.1.00995.S, and 2019.1.01559.S (PI: Murchikova). 
All observations are conducted at around 230 GHz and capture the \bco\ line at 230.538 GHz.
Every observation used J1924-2914 ($\sim$2~Jy) for flux/bandpass calibration and J1744-3116 (0.27$\pm$0.03~Jy) for complex gain calibration.
We adopt ALMA's standard Band 6 calibration uncertainty of 10\% on the flux values.
The data have a half-power primary beam width (i.e., limitation of the telescope's field of view due to the size of 12 m antennas) of 22.7\arcsec \, at 230.538~GHz.

Calibration of the \bco\ data was conducted in several steps.
{\sgra\ varies on timescales of seconds and the flux can change by $\sim1$~Jy within half hour (e.g., \citealt{Murchikova2021}).}
We used \verb|UVmultifit| \citep{marti-vidal2014} to fit the variability of \sgra \, at uv-level.
We then used the time-dependent lightcurve to self-calibrate the data.
Finally, we removed the continuum emission of the black hole from the data at the uv-level to improve the quality of imaging.
The imaging was conducted with the Common Astronomy Software Applications (CASA) using task \verb|tclean| with multiscale deconvolution (scales: 0, 7, 42, 126 pixels; and Briggs weighting: 0.5).
The primary beam's first null is at 27.1\arcsec \, from the center of the image.
The produced image has 0.02\arcsec pixels and a diameter of about 54\arcsec\ or $\sim$2~pc at the approximate distance to the Galactic Center of $D=8.3$~kpc.
The produced image covers the primary beam nearly completely.

\begin{figure}
    \centering
    \includegraphics[width=\linewidth]{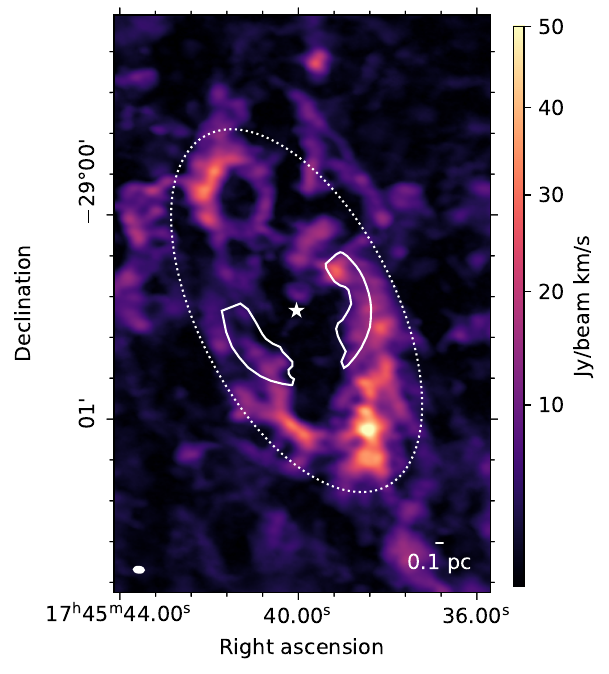}
    \caption{
    The full map of the CND in \ahcn\ emission from  \citet{Hsieh2021}. 
    The dotted ellipse shows the extent of the CND \citep{Tsuboi2018}. 
    The part of the CND considered for the calculations of the CO--to--\mh\ conversion factor is outlined with the white contours.
    The position of \sgra\ is marked with a white star.
    }
    \label{fig:CNDwhole}
\end{figure}

The absolute shortest projected baseline in the observations has a length of 12.4~m for a maximum recovered scale of 26.4\arcsec.
\bco\ data cubes have 25~$\mathrm{km \, s}^{-1}$ spectral resolution and 80 $\mu$Jy rms noise, smoothed to a common beam size of 0.256\arcsec×0.235\arcsec \, with a postilion angle of 60.3\deg. 
{Integrated intensity maps are made using pixels with values $>5\sigma$.}
Velocities are in the LSRK frame (radio convention).
Continuum subtraction of Sgr~A~West was done in the image domain via linear fitting to line-free channels.
The inner 0.5\arcsec \, (0.002 pc) radius around the \sgra \, is masked due to residual (i.e., not fully removed at the uv-level) black hole emission.

{We also adopt \ahcn\ data cubes from the dataset of \citealt{Hsieh2021}. \ahcn\ data is included in their ${\rm CS}(J=2\to1)$ observations (ALMA project 2017.1.00040.S).
The synthesized beam for \ahcn\ data is 3.69\arcsec×2.33\arcsec \ with a position angle of 85.1\deg, the cube has an rms of 0.11~Jy/beam, and a spectral resolution of 0.8~\kms.}

\subsection{Flux recovery}\label{ss:recoveredflux}

An incomplete sampling of the uv-plane particularly at short spacings, can lead to resolved-out flux.
To mitigate this issue, we (i) restrict our flux measurements to regions that are well resolved (see Figure \ref{fig:CNDreg} white contours); and (ii) compare the recovered flux to \aco\ measurements with the Nobeyama 45~m single-dish telescope \citep{Tokuyama2019}, assuming a $\bco / \aco$ ratio of 0.85 with an uncertainty of $\pm$0.2.
We adopt this value because it is the mean between the optically thick limit (a ratio of 1) and the value of 0.7 that is commonly measured in the local universe \citep{Sandstrom2013}.
The uncertainty of $\pm$0.2 includes both conditions. 
{We estimate the filling factor from the maximum linear size (34\arcsec) of the individual regions captured within the white contours in Figure \ref{fig:CNDreg}.
The resolution of the Nobeyama data is 15\arcsec.
The filling factor is thus the ratio of the an individual area in Figure \ref{fig:CNDreg} and the 15\arcsec$\times$34\arcsec\ region sampled by Nobeyama.
The respective values for the west side and east side are  0.436 and 0.444.
The expected \bco\ integrated flux density would be 14,300$\pm$3000~Jy~\kms. 
We find that the flux recovered in our ALMA observations is  $88\pm21$\% of the flux observed by the single-dish Nobeyama 45~m telescope and thus consistent with all flux being recovered.}

\section{CO--to--\mh\ conversion factor in Milky Way's Galactic Center} \label{sec:co2mh}

\subsection{Caculating by area}\label{ss:byarea}

Consider the total integrated \bco\ emission of the CND {captured within the white contour} on our map (Figure \ref{fig:CNDreg}).
The total integrated flux in this region is {12,400~Jy~\kms}.
The area of the CND captured in the image and used for calculation is 460.6 $\text{arcsec}^{2}.$ 
As the model for the full CND, we adopt an ellipse with a semi-major axis of 58\arcsec, an inclination of 30\deg, and the area of 5284.2 $\text{arcsec}^{2}$ (dotted line in Figure \ref{fig:CNDwhole}, \citealp{Tsuboi2018}). 
The fraction of the full CND captured in \bco \ map (see white contours in Figure \ref{fig:CNDwhole}) is about 8.7\%.
The total molecular mass of the CND is usually estimated to be a few $10^4$~\Msun \, but the exact value varies \citep{James2021, Hsieh2018, Tsuboi2018, Requena-Torres2012, Lau2013, Liu2013,Oka2011, Etxaluze2011}. Here we adopt the value of $3.0 \pm 1.5\times10^4$~\Msun, which is closest to the estimate of \citealt{Tsuboi2018}. 
In the calculation of the CO-to-\mh\ conversion factor below, the errorbars are dominated by the uncertainties on the mass of the CND.

Here we use a purely geometrical approach and assume that the fraction of the captured area corresponds to the fraction of mass in this region. 
So we expect about 8.7\% of the mass of the CND to be within the selected regions in Figure \ref{fig:CNDreg}. 
This gives the value of the CO-to-\mh \ conversion factor at 
\begin{eqnarray*}
\alpha_{\rm{CO}}\big|_\mathrm{by \, area }&\simeq& 4.3\pm2.3 \, M_\odot (\mathrm{K \, km \, s^{-1} \, pc^2})^{-1}.
\end{eqnarray*}

This approach does not take into account density variations in the CND. 
We consider density variations below.

\subsection{Calculating by fraction of HCN emission}

Let us assume that the density variations in the CND are relatively well represented by the variations of the integrated \ahcn\ flux, and calculate the ratio of the integrated \ahcn\ flux within selected regions on Figure \ref{fig:CNDreg} to the whole CND. We use \ahcn\ as it is probably the best studied CND map \cite{Hsieh2021}. 

The total \ahcn\ integrated flux density  of the CND is 3700~Jy~\kms (Figure \ref{fig:CNDwhole}). 
The fraction of \ahcn\ emission captured within the white contours in Figures \ref{fig:CNDreg}-\ref{fig:CNDwhole} is about 350$\pm$35~Jy~\kms, i.e., 9.7\%, implying that H$_2$ mass in this region is $2.9\pm1.4\times10^3~M_\odot$.
So this gives the value of the CO-to-\mh \ conversion factor at
\begin{eqnarray*}
\alpha_{\rm{CO}}\big|_\mathrm{by \, HCN \, fraction }&\simeq& 4.7\pm2.6 \, M_\odot (\mathrm{K \, km \, s^{-1} \, pc^2})^{-1}.
\end{eqnarray*}
We see that calculations of the $\alpha_{\rm{CO}}$ conversion factor by the captured area or by the fraction of the \ahcn\ emission of the CND give nearly the same result.

\section{Foreground Absorption}

Line-of-sight absorption towards the Galactic Center has a potential to effect the calculations of the of $\alpha$[CO] and X[CO]. 
The absorption is significant  at velocities of -150 to -30~\kms\ for the southern lobe of the CND, and 80 to 150~\kms\ for the northern lobe of the CND, as a result of the presence of the Galactic spiral arms and other foreground \citep[e.g.,][]{Christopher2005,Requena-Torres2012}.

To estimate the absorption, we use the radiative transfer code RADEX \citep{vanderTak2007} to estimate the expected brightness temperature of the CND, $T_0$, from its known average properties. The CND has an average temperature of 100~K, CO column density of 10$^{18}$~\cc, \mh\ density of 10$^5$~\ccc, and line width of 15~\kms\ \citep{Hsieh2021,Requena-Torres2012}, resulting  in an estimated average ${T_0}=$700 K~\kms.
The observed integrated intensity (Figure \ref{fig:CNDreg}) in radio units is 695K~\kms.

For the foreground, we assume there are six absorption features from the spiral arms and other orbits crossing the line-of-sight toward the Galactic center, with an average brightness temperature of $T_\mathrm{foreground} = 29$~K~\kms.
For RADEX, we assume an average \mh\ density of 10$^3$~\ccc, velocity widths of $\sim$20~\kms, temperatures of 10~K, and column CO densities 10$^{17}$~\cc\ per feature, resulting in the total column density of $6\times10^{17}\cc$ \citep{Bieging1980,Binney1991,Liszt1978,Scoville1972,Oort1977,Menon1970}.

To estimate the absorption, we use the radiative transfer formula
\begin{eqnarray*}
    \rm{T_{obs}= T_{foreground}(1-e^{-\tau}) + T_0(e^{-\tau})}.
\end{eqnarray*}
Here $\rm{T_{obs}}$ is the observed integrated intensity, $\rm{T_{foreground}}$ is the foreground integrated intensity, $\rm{T_0}$ is the integrated intensity of the CND, and $\tau$ is the optical depth of the foreground obscuring material.

We obtain $\tau\simeq0.007$ and the foreground absorption of just $\lesssim1\%.$
This is not surprising, as the most significant absorption affects -135, -50, and -30~\kms\ features of the blue shifted lobe of the CND \citep{Christopher2005,Requena-Torres2012}, and
by selecting a sub-region near the middle of the CND, we avoid the most significant effects.

\section{Summary} \label{sec:assume}

We have provided the first estimate of the CO--to--\mh\ conversion factor in the 1~pc scale nuclear region of the Milky Way. 
We calculated the $\alpha_{\rm{CO}}$ value in two ways, which produced close values. 
We report the average as the final result: 
\begin{eqnarray*}
\alpha_{\rm{CO}}&\simeq& 4.5\pm2.5 \, M_\odot (\mathrm{K \, km \, s^{-1} \, pc^2})^{-1}, 
\end{eqnarray*}
which is equivalent to 
\begin{eqnarray*}
\mathrm{X[CO]}^{} &\simeq& 2.1\pm1.1\times 10^{20} \, \mathrm{cm}^{-2} (\mathrm{K \, km \, s^{-1}})^{-1}.
\end{eqnarray*}

The value of the CO--to--\mh\ conversion factor in the inner $\sim 1$ pc of the black hole \sgra \, {found  is to be slightly higher but consistent within the uncertainties with that found for the disk of the Milky Way $\alpha_{\rm{CO}} \big|_{\mathrm{MW\,disk}}=4.3\pm1.3$~\Msun(\Lkkmspc)$^{-1}.$
However, it is higher than what is known of the larger central molecular zone $\alpha_{\rm{CO}} \big|_{\mathrm{MW \,CMZ}}=1.0\pm0.3~M_\odot (\mathrm{K \, km \, s^{-1} \, pc^2})^{-1} $ \citep{Kohno2024}.
This suggests that the CND may not inherit the properties of the larger CMZ.
The amount of molecular gas is, of course, expected to greatly decrease closer to the black hole.  
The uncertainties are large due to the wide variations in the estimated mass of the molecular gas in the Galactic Center's CND, on which we rely in our calculations.
However, it is also possible, if the optical depth of the circumnuclear disk (CND) for the \bco/\aco \, emission is on the order of a few, then the observed \bco/\aco \ emission originates primarily from the surface of the CND rather than probing its full depth. This would lead to an overestimation of the CO--to--\mh\ conversion factor by a similar factor as the optical depth.
}

\begin{acknowledgments}

We are grateful to Claire J. Chandler for advice on data processing, and to Ivan Marti-Vidal, Anna Ciurlo, Mark Morris, Shoko Sakai, Tuan Do, Roger Blandford, Nick Scoville, Rick Perley, and Elizabeth Mills for comments and discussions, and to Jurgen Ott, Jin Koda, Eliot Quataert, and Sean Ressler for contribution to writing ALMA proposals, and to the anonymous referee for their helpful comments.

M.D.G. is supported by the CIERA Postdoctoral Fellowship from the Center for Interdisciplinary Exploration and Research in Astrophysics at Northwestern University.

This paper makes use of the following ALMA data: ADS/JAO.ALMA \#2016.1.00870.S, \#2017.1.00995.S, \#2017.1.00040.S, and \#2019.1.01559.S.
ALMA is a partnership of ESO (representing its member states), NSF (USA) and NINS (Japan), together with NRC (Canada) and NSC and ASIAA (Taiwan) and KASI (Republic of Korea), in cooperation with the Republic of Chile. 
The Joint ALMA Observatory is  operated by ESO, AUI/NRAO and NAOJ.

The National Radio Astronomy Observatory is a facility of the National Science Foundation operated under cooperative agreement by Associated Universities, Inc.

This work used computing resources provided by Northwestern University and the Center for Interdisciplinary Exploration and Research in Astrophysics (CIERA). This research was supported in part through the computational resources and staff contributions provided for the Quest high performance computing facility at Northwestern University  which is jointly supported by the Office of the Provost, the Office for Research, and Northwestern University Information Technology.
  
\end{acknowledgments}

\facilities{ALMA}

\software{astropy \citep{Astropy2013,Astropy2018,Astropy2022},  
          UVmultifit \citep{marti-vidal2014}, 
          CASA \citep{CASATeam2022PASP}
          }



\bibliography{bib_alphaCO,bib_sgra,bib_soft,bib_misc}
\bibliographystyle{aasjournalv7}



\end{document}